\documentclass[10pt,a4paper]{article}

\usepackage[utf8]{inputenc}
\usepackage[english]{babel}

\begin{document}
\large

\newpage
\begin{center}
\LARGE
{\bf A New Family with a Fourth Lepton Flavor}
\end{center}
\large
\vspace{0.1cm}
\begin{center}
{\bf Rasulkhozha S. Sharafiddinov}
\end{center}
\vspace{0.1cm}
\begin{center}
{\bf Institute of Nuclear Physics, Uzbekistan Academy of Sciences,
\\Tashkent, 100214 Ulugbek, Uzbekistan}
\end{center}
\vspace{0.1cm}

\begin{center}
{\bf Abstract}
\end{center}

We present here arguments in favor of the existence of the lightest lepton and its neutrino. 
This new family with a fourth lepton flavor uncovered so far unobserved universal properties of matter. The unity of the laws of their nature admits the flavor symmetrical decays of the electron and proton. Thereby, it predicts the new modes in the decays of the muon, tau lepton and neutron. 
At the same time, in all these transitions, no conservation laws are violated, confirming the availability in nature of a place for an additivity of mass.
 
\vspace{0.8cm}
\noindent
{\bf 1. Introduction}
\vspace{0.4cm}

The unity of symmetry laws establishes in nature a highly important connection between the 
different spin states of each of all types of neutrinos. Such a principle, regardless of what 
is the mechanism of a flavor symmetrical mode of neutrino oscillation [1], must be considered 
as a criterion for unification of fermions at the new level [2] that to any type of the left (right)-handed lepton corresponds a kind of neutrino. This gives the possibility to define [3,4] 
the family structure of leptons in the presence not only of the left-handed but also of the 
right-handed doublets. They may be written as
\begin{equation}
..., \pmatrix{\nu_{e}\cr e^{-}}_{L},
(\nu_{e}, \, \, \, \, e^{-})_{R}, \, \, \, \,
\pmatrix{\nu_{\mu}\cr \mu^{-}}_{L},
(\nu_{\mu}, \, \, \, \, \mu^{-})_{R}, \, \, \, \,
\pmatrix{\nu_{\tau}\cr \tau^{-}}_{L},
(\nu_{\tau}, \, \, \, \, \tau^{-})_{R}, ...,
\label{1}
\end{equation}
\begin{equation}
..., \pmatrix{{\bar \nu_{e}}\cr e^{+}}_{R},
({\bar \nu_{e}}, \, \, \, \, e^{+})_{L}, \, \, \, \,
\pmatrix{{\bar \nu_{\mu}}\cr \mu^{+}}_{R},
({\bar \nu_{\mu}}, \, \, \, \, \mu^{+})_{L}, \, \, \, \,
\pmatrix{{\bar \nu_{\tau}}\cr \tau^{+}}_{R},
({\bar \nu_{\tau}}, \, \, \, \, \tau^{+})_{L}, ....
\label{2}
\end{equation}

Such a presentation is based logically on the question as to what is the value of masses 
of so far unknown [5,6] families of leptons.

Each family of the known types of leptons has [7,8] his own flavor. Therefore, it should at a given stage be characterized any particle by the three $(l=e,$ $\mu,$ $\tau,...)$ lepton flavors:
\begin{equation}
L_{l}=\left\{
{\begin{array}{l}
{+1\quad \mbox{for}\quad l^{-}_{L}, \, \, \, \, \, l^{-}_{R}, \, \, \, \,
\nu_{lL}, \, \, \, \, \nu_{lR},}\\
{-1\quad \mbox{for}\quad l^{+}_{R}, \, \, \, \, l^{+}_{L}, \, \, \, \,
{\bar \nu_{lR}}, \, \, \, \, {\bar \nu_{lL}},}\\
{\, \, \, \, \, 0\quad \mbox{for}\quad \mbox{remaining particles.}}\\
\end{array}}\right.
\label{3}
\end{equation}

Conservation both of full lepton number
\begin{equation}
L_{e}+L_{\mu}+L_{\tau}=const
\label{4}
\end{equation}
and of all forms of flavors
\begin{equation}
L_{l}=const
\label{5}
\end{equation}
is, by itself, not excluded [9]. This indicates that the decays
\begin{equation}
\mu^{-}_{L,R}\rightarrow e^{-}_{L,R}{\bar \nu_{e R,L}}\nu_{\mu L,R}, \, \, \, \,
\mu^{+}_{R,L}\rightarrow e^{+}_{R,L}\nu_{e L,R}{\bar \nu_{\mu R,L}},
\label{6}
\end{equation}
\begin{equation}
\tau^{-}_{L,R}\rightarrow e^{-}_{L,R}{\bar \nu_{e R,L}}\nu_{\tau L,R}, \, \, \, \,
\tau^{+}_{R,L}\rightarrow e^{+}_{R,L}\nu_{e L,R}{\bar \nu_{\tau R,L}},
\label{7}
\end{equation}
\begin{equation}
\tau^{-}_{L,R}\rightarrow \mu^{-}_{L,R}{\bar \nu_{\mu R,L}}\nu_{\tau L,R}, \, \, \, \,
\tau^{+}_{R,L}\rightarrow \mu^{+}_{R,L}\nu_{\mu L,R}{\bar \nu_{\tau R,L}}
\label{8}
\end{equation}
and other transitions can carry out in leptons at the emission of difermions [3,4] 
of a definite flavor 
\begin{equation}
(e_{L}^{-}, {\bar \nu_{eR}}), \, \, \, \,
(e_{R}^{-}, {\bar \nu_{eL}}),
\label{9}
\end{equation}
\begin{equation}
(e_{R}^{+}, \nu_{eL}), \, \, \, \,
(e_{L}^{+}, \nu_{eR}),
\label{10}
\end{equation}
\begin{equation}
(\mu_{L}^{-}, {\bar \nu_{\mu R}}), \, \, \, \,
(\mu_{R}^{-}, {\bar \nu_{\mu L}}),
\label{11}
\end{equation}
\begin{equation}
(\mu_{R}^{+}, \nu_{\mu L}), \, \, \, \,
(\mu_{L}^{+}, \nu_{\mu R}).
\label{12}
\end{equation}

Each paraparticle here is responsible for conservation of summed electric charge. Thereby, it expresses the idea of charge quantization law. 

According to one of its dynamical aspects, the value [10,11] of lepton electric charge $e_{l}^{E}$ is connected with some universal charge $e_{0}^{E}$ and equal to
\begin{equation}
e_{l}^{E}=ne_{0}^{E}, \, \, \, \, n=0,1,2,...
\label{13}
\end{equation}
in which $e_{0}^{E}$ was accepted as an electron charge.

It is interesting, however, that the electron, muon and tau lepton having an equal charge 
possess the different masses. On the other hand, as noted in the work [12] for the fist 
time, a mass spectrum of elementary particles must be restricted from above and below 
by masses of the limited size.

If we now take into account that the difference in masses of leptons is of the unified 
principle [13], then there arises a question of whether the acceptance of fundamental charge $e_{0}^{E}$ in Eq. (\ref{13}) as an electron charge is not strictly nonverisimilar even at 
its universality properties.

It is also relevant to recall the mass-charge duality [14], according to which, any of all types 
of charges may serve as a certain indication to the existence in nature of a kind of inertial mass. 
The mass and charge of an electroweakly charged lepton are naturally united in rest mass $m_{l}$
and charge $e_{l}$ equal to the electroweak $(EW)$ mass and charge
\begin{equation}
m_{l}=m_{l}^{EW}=m_{l}^{E}+m_{l}^{W},
\label{14}
\end{equation}
\begin{equation}
e_{l}=e_{l}^{EW}=e_{l}^{E}+e_{l}^{W}
\label{15}
\end{equation}
consisting of weak $(W)$ and Coulomb $(E)$ parts.

Therefore, without contradicting ideas of charge quantization, we conclude that regardless 
of what are the maximally and minimally possible values of the electric and weak types of masses, the crossing of their spectra corresponds in them to the existence of hitherto unobserved particle
with some universal mass.

Our purpose in a given work is to elucidate whether there exists any new family of leptons and, if so, what the expected fermions say about so far unknown properties of matter.

\vspace{0.8cm}
\noindent
{\bf 2. From earlier symmetries to new leptons}
\vspace{0.4cm}

The nature relates each part of the unified mass of a particle to corresponding contribution of 
the structural components of its united charge. In the limit of electroweakly charged leptons, 
these connections [15] have the form
\begin{equation}
e_{l}^{E}=-g_{V_{l}}\frac{m_{l}^{E}m_{l}^{W}}
{2m_{W}^{2}}\frac{1}{\sin\theta_{W}},
\label{16}
\end{equation}
where $e_{l^{-}}^{E}$ and $e_{l^{+}}^{E}$ distinguish from one another by a sign, $m_{W}$ denotes the $W^{\pm}$-boson mass, and a constant $g_{V_{l}}=2\sin^{2}\theta_{W}-(1/2)$ characterizes 
the leptonic $(l=l,$ $\nu_{l})$ weak current vector part.

Universality of the size of $|e_{l}^{E}|,$ describing the fact that a multiplier $m_{l}^{E}m_{l}^{W}$ must in Eq. (\ref{16}) be one of fundamental physical parameters,
leads us to a lepton  universality principle [15] that
\begin{equation}
m_{l}^{E}m_{l}^{W}=const.
\label{17}
\end{equation}

The absence of one of $m_{l}^{E}$ or $m_{l}^{W},$ as we see, would imply that neither exists at all. From such a point of view, earlier experiments [16] about electric masses of leptons
\begin{equation}
m_{e}^{E}=0.51\ {\rm MeV},
\label{18}
\end{equation}
\begin{equation}
m_{\mu}^{E}=105.658\ {\rm MeV},
\label{19}
\end{equation}
\begin{equation}
m_{\tau}^{E}=1776.99\ {\rm MeV}
\label{20}
\end{equation}
may serve as the first source of facts predicting their weak masses
\begin{equation}
m_{e}^{W}=5.15\cdot 10^{-2}\ {\rm eV},
\label{21}
\end{equation}
\begin{equation}
m_{\mu}^{W}=2.49\cdot 10^{-4}\ {\rm eV},
\label{22}
\end{equation}
\begin{equation}
m_{\tau}^{W}=1.48\cdot 10^{-5}\ {\rm eV}.
\label{23}
\end{equation}

The basis for our choice is that the current used in the devices for measurements of $m_{l}$ 
and $e_{l}$ has a Coulomb nature.

Masses of both types satisfy in addition a constancy of the size
\begin{equation}
m_{l}^{E}m_{l}^{W}=26318.11\ {\rm eV^{2}}
\label{24}
\end{equation}
connected with a latent regularity of nature of spectra of lepton EW masses.

There exist, however, the maximally $m_{lmax}^{K}$ and minimally $m_{lmin}^{K}$ possible limits 
both on the electric $(K=E)$ and on the weak $(K=W)$ masses of leptons in the spectra of 
elementary particle masses.

Furthermore, if it turns out that $m_{l}^{E}\rightarrow m_{lmax}^{E}$ implies
$m_{l}^{W}\rightarrow m_{lmin}^{W},$ and $m_{l}^{W}\rightarrow m_{lmax}^{W}$
says $m_{l}^{E}\rightarrow m_{lmin}^{E},$ comparing Eqs. (\ref{18})-(\ref{20})
with the corresponding masses from Eq. (\ref{21}) to Eq. (\ref{23}), it is easy to
see a coincidence of their spectra
\begin{equation}
m_{l}^{E}=m_{l}^{W},
\label{25}
\end{equation}
which takes place only at
\begin{equation}
m_{l}^{E}\ll m_{e}^{E}, \, \, \, \,
m_{l}^{W}\gg m_{e}^{W}.
\label{26}
\end{equation}

This property corresponds in nature to the same type of charged lepton $(l=\epsilon)$ possessing 
the universal mass. It can be called an evrmion. We use in addition the symbols $\epsilon^{-}$ 
and $\epsilon^{+}$ denoting the evrmion and its antiparticle.

One of highly important features of the suggested lepton is the universality of the square of 
any of the structural parts of its mass. Indeed, uniting Eq. (\ref{17}) with Eq. (\ref{25}), 
we establish a connection
\begin{equation}
(m_{\epsilon}^{K})^{2}=m_{\epsilon}^{E}m_{\epsilon}^{W}=
m_{l}^{E}m_{l}^{W}=const.
\label{27}
\end{equation}

Thus, earlier experiments [17] about lepton universality [18-21] may testify in favor of the existence of the lightest lepton, namely, of the evrmion having the electric mass and charge
\begin{equation}
m_{\epsilon}^{E}=162,22857\ {\rm eV},
\label{28}
\end{equation}
\begin{equation}
e_{\epsilon}^{E}=1.602\cdot 10^{-19}\ {\rm C}.
\label{29}
\end{equation}

Therefore, it is not surprising that if Eq. (\ref{27}) does not change, the mass and charge 
of the evrmion play a role of units of mass and charge
\begin{equation}
m_{0}^{E}=m_{\epsilon}^{E}, \, \, \, \, e_{0}^{E}=e_{\epsilon}^{E}.
\label{30}
\end{equation}

Of course, our definition of fundamental constants (\ref{30}) is not a standard one. This, however,
requires one to follow the logic of a unified nature of mass and charge from the point of view of 
mass-charge duality.

\vspace{0.8cm}
\noindent
{\bf 3. Earlier symmetries about new neutrinos}
\vspace{0.4cm}

We see that nature itself is not in force to define the same components of mass and charge
regardless of their other parts. It relates herewith each component of mass even in the case 
of a neutrino to corresponding part of its charge.

Furthermore, if neutrinos are of families of the studied types of leptons, this connection [22] appears as follows:
\begin{equation}
e_{\nu_{l}}^{E}=-g_{V_{\nu_{l}}}\frac{m_{\nu_{l}}^{E}m_{\nu_{l}}^{W}}
{2m_{W}^{2}}\frac{1}{\sin\theta_{W}}.
\label{31}
\end{equation}
Here, $e_{\nu_{l}}^{E}$ for the neutrino (antineutrino) has the negative
(positive) sign.

Comparison of Eq. (\ref{31}) with Eq. (\ref{16}) transforms it to the form
\begin{equation}
e_{\nu_{l}}^{E}=\frac{m_{\nu_{l}}^{E}m_{\nu_{l}}^{W}}
{m_{e}^{E}m_{e}^{W}}e_{e}^{E}.
\label{32}
\end{equation}

Such a connection would seem to exist [23,24] only at the absence of gauge invariance. But 
unlike the earlier descriptions of nature of this symmetry, its mass structure [25] establishes 
Eq. (\ref{32}) from the point of view of a unified principle.

We recognize that Eq. (\ref{32}) is incompatible with charge [10,11] quantization, if it 
does not possess any structural properties. On the other hand, as stated in Eq. (\ref{32}), each electrically charged particle says in favor of a kind of magnetically charged monoparticle [26]. 
In such a situation, the same neutrino can lead to quantization of magnetic charges of all mononeutrinos and vice versa.

In a similar way, one can as an example introduce an arbitrary electric charge [27]. At the same time, the standard electroweak theory [28-30], by itself, does not exclude a given procedure.

It is clear, however, that the available laboratory facts [16] define only the upper limits of the 
neutrino mass and charge. There are also the earlier [31] and comparatively new [32] experiments, the analysis of which predicts the existence of neutrino universality. A beautiful example [32] 
is the following:
\begin{equation}
e_{\nu_{l}}^{E}< 2\cdot 10^{-13}\ {\rm e_{e}^{E}}.
\label{33}
\end{equation}

The solutions (\ref{17}) and (\ref{32}) indicate herewith to a principle that
\begin{equation}
m_{\nu_{l}}^{E}m_{\nu_{l}}^{W}=const.
\label{34}
\end{equation}

To the same implication, one can also lead by another way starting from the full lepton number, 
the conservation of which [22] establishes a connection
\begin{equation}
m_{\nu_{e}}^{E}m_{\nu_{e}}^{W}:m_{\nu_{\mu}}^{E}m_{\nu_{\mu}}^{W}:
m_{\nu_{\tau}}^{E}m_{\nu_{\tau}}^{W}=
m_{e}^{E}m_{e}^{W}:m_{\mu}^{E}m_{\mu}^{W}:m_{\tau}^{E}m_{\tau}^{W}.
\label{35}
\end{equation}

Unification of Eq. (\ref{17}) with Eq. (\ref{35}) leads us to Eq. (\ref{34}) once more, confirming the availability in it of a sign of an equality.

One of masses $m_{\nu_{l}}^{E}$ or $m_{\nu_{l}}^{W},$ as noted in Eq. (\ref{34}), can exist only in the presence of the second of them. From such a point of view, known [16] experimental restrictions on the electric masses
\begin{equation}
m_{\nu_{e}}^{E}< 2.5\ {\rm eV},
\label{36}
\end{equation}
\begin{equation}
m_{\nu_{\mu}}^{E}< 0.17\ {\rm MeV},
\label{37}
\end{equation}
\begin{equation}
m_{\nu_{\tau}}^{E}< 18.2\ {\rm MeV}
\label{38}
\end{equation}
suggest the measured for the first time sizes of neutrino weak masses
\begin{equation}
m_{\nu_{e}}^{W}< 2.1\cdot 10^{-9}\ {\rm eV},
\label{39}
\end{equation}
\begin{equation}
m_{\nu_{\mu}}^{W}< 3.096\cdot 10^{-14}\ {\rm eV},
\label{40}
\end{equation}
\begin{equation}
m_{\nu_{\tau}}^{W}< 2.89\cdot 10^{-16}\ {\rm eV}.
\label{41}
\end{equation}

Their comparison gives the right to state that a constancy of multiplier
\begin{equation}
m_{\nu_{l}}^{E}m_{\nu_{l}}^{W}< 52636.22\cdot 10^{-13}\ {\rm eV^{2}}
\label{42}
\end{equation}
is connected with its structure depending on nature of spectra of neutrino EW masses.

They show that the existence in the spectra of lepton masses both of the maximally $m_{\nu_{l}max}^{K}$ and of the minimally $m_{\nu_{l}min}^{K}$ limited values of electric 
and weak masses of neutrinos is practically not excluded.

Uniting Eqs. (\ref{36})-(\ref{38}) with Eqs. (\ref{39})-(\ref{41}) and by following
that $m_{\nu_{l}}^{E}\rightarrow m_{\nu_{l}max}^{E}$ expresses
$m_{\nu_{l}}^{W}\rightarrow m_{\nu_{l}min}^{W},$ and
$m_{\nu_{l}}^{W}\rightarrow m_{\nu_{l}max}^{W}$ describes
$m_{\nu_{l}}^{E}\rightarrow m_{\nu_{l}min}^{E},$ we observe
the crossing of spectra of both types of neutrino masses
\begin{equation}
m_{\nu_{l}}^{E}=m_{\nu_{l}}^{W},
\label{43}
\end{equation}
which is realized only if
\begin{equation}
m_{\nu_{l}}^{E}\ll m_{\nu_{e}}^{E}, \, \, \, \,
m_{\nu_{l}}^{W}\gg m_{\nu_{e}}^{W}.
\label{44}
\end{equation}

This picture has important consequences for the same type of neutrino corresponding in nature 
to the evrmion. Therefore, from its point of view, it should be expected that the square of any of the structural components of mass of the suggested neutrino possesses the universality properties. Indeed, starting from Eqs. (\ref{34}) and (\ref{43}), we are led to a principle that
\begin{equation}
(m_{\nu_{\epsilon}}^{K})^{2}=
m_{\nu_{\epsilon}}^{E}m_{\nu_{\epsilon}}^{W}=
m_{\nu_{l}}^{E}m_{\nu_{l}}^{W}=const.
\label{45}
\end{equation}

The latter together with Eq. (\ref{42}) convinces us here that the fact [31,32] of neutrino universality is the first confirmation of the existence of the lightest neutrino, namely, 
of the evrmionic neutrino having the electric charge (\ref{33}) and mass
\begin{equation}
m_{\nu_{\epsilon}}^{E}< 7.2550823\cdot 10^{-5}\ {\rm eV}.
\label{46}
\end{equation}

It is not excluded, however, that if Eq. (\ref{45}) holds, the mass and charge of the evrmionic neutrino refer doubtless only to fundamental physical parameters. They of course characterize 
those particles, the charge of which is not comparable with an evrmion charge.

\vspace{0.8cm}
\noindent
{\bf 4. The fourth lepton flavor}
\vspace{0.4cm}

Each of leptons has his own neutrino. If such a pair is the evrmion and its neutrino, they constitute the new leptonic families.

For elucidation of their ideas, it is desirable to present the family structure of leptons 
(\ref{1}) and (\ref{2}) in the general form by the following manner:
$$\pmatrix{\nu_{\epsilon}\cr \epsilon^{-}}_{L},
(\nu_{\epsilon}, \, \, \, \, \epsilon^{-})_{R}, \, \, \, \,
\pmatrix{\nu_{e}\cr e^{-}}_{L},
(\nu_{e}, \, \, \, \, e^{-})_{R},$$
\begin{equation}
\pmatrix{\nu_{\mu}\cr \mu^{-}}_{L},
(\nu_{\mu}, \, \, \, \, \mu^{-})_{R}, \, \, \, \,
\pmatrix{\nu_{\tau}\cr \tau^{-}}_{L},
(\nu_{\tau}, \, \, \, \, \tau^{-})_{R}, ...,
\label{47}
\end{equation}
$$\pmatrix{{\bar \nu_{\epsilon}}\cr \epsilon^{+}}_{R},
({\bar \nu_{\epsilon}}, \, \, \, \, \epsilon^{+})_{L}, \, \, \, \,
\pmatrix{{\bar \nu_{e}}\cr e^{+}}_{R},
({\bar \nu_{e}}, \, \, \, \, e^{+})_{L},$$
\begin{equation}
\pmatrix{{\bar \nu_{\mu}}\cr \mu^{+}}_{R},
({\bar \nu_{\mu}}, \, \, \, \, \mu^{+})_{L}, \, \, \, \,
\pmatrix{{\bar \nu_{\tau}}\cr \tau^{+}}_{R},
({\bar \nu_{\tau}}, \, \, \, \, \tau^{+})_{L}, ....
\label{48}
\end{equation}

Any evrmionic family here must distinguish itself from earlier leptons by the individual flavor. This new lepton flavor $(L_{\epsilon})$ can be called an evrmion number. We have, thus, a real possibility to characterize each particle by the four $(l=\epsilon,$ $e,$ $\mu,$ $\tau,...)$ 
lepton (\ref{3}) flavors.

Conservation of full lepton number
\begin{equation}
L_{\epsilon}+L_{e}+L_{\mu}+L_{\tau}=const
\label{49}
\end{equation}
and all types of lepton flavors (\ref{5}) indicates to the existence in nature of the most
diverse connections with evrmions and their neutrinos.

\vspace{0.8cm}
\noindent
{\bf 5. Mass criterion for the electron decay}
\vspace{0.4cm}

If the light fermions having a fourth lepton flavor exist, they must birth in the decays 
of the more heavy particles. An example of this may be a new scheme of the muon decay
\begin{equation}
\mu^{-}_{L,R}\rightarrow \epsilon^{-}_{L,R}{\bar \nu_{\epsilon R,L}}\nu_{\mu L,R}, \, \, \, \,
\mu^{+}_{R,L}\rightarrow \epsilon^{+}_{R,L}\nu_{\epsilon L,R}{\bar \nu_{\mu R,L}}. 
\label{50}
\end{equation}

Its legality follows from the unified force that unites the two left (right)-handed particles 
of the suggested evrmionic family in dievrmions
\begin{equation}
(\epsilon_{L}^{-}, {\bar \nu_{\epsilon R}}), \, \, \, \,
(\epsilon_{R}^{-}, {\bar \nu_{\epsilon L}}),
\label{51}
\end{equation}
\begin{equation}
(\epsilon_{R}^{+}, \nu_{\epsilon L}), \, \, \, \,
(\epsilon_{L}^{+}, \nu_{\epsilon R}).
\label{52}
\end{equation}

These parafermions are in favor of conservation of full lepton number and all four types of lepton flavors admitting the existence of the new scheme in the decay of the $\tau$-lepton
\begin{equation}
\tau^{-}_{L,R}\rightarrow \epsilon^{-}_{L,R}{\bar \nu_{\epsilon R,L}}\nu_{\tau L,R}, \, \, \, \,
\tau^{+}_{R,L}\rightarrow \epsilon^{+}_{R,L}\nu_{\epsilon L,R}{\bar \nu_{\tau R,L}}.
\label{53}
\end{equation}

We have already seen that any of the decays (\ref{50}) and (\ref{53}) is carried out in nature 
by the same criterion. Such a criterion can, for example, be a difference in masses of leptons. It establishes not only a lepton or a neutrino universality but also a flavor symmetrical connection,
at which the birth of dievrmions (\ref{51}) and (\ref{52}) originates in the transitions of one heavy lepton into another state with the lowest mass. In a given case, a new example 
of the $\beta$-transition is the electron decay
\begin{equation}
e^{-}_{L,R}\rightarrow \epsilon^{-}_{L,R}{\bar \nu_{\epsilon R,L}}\nu_{e L,R}, \, \, \, \,
e^{+}_{R,L}\rightarrow \epsilon^{+}_{R,L}\nu_{\epsilon L,R}{\bar \nu_{e R,L}}.
\label{54}
\end{equation}

Many authors state that an electron must not decay. Its decay [33,34] would seem to contradict our observation that stable electrons constitute the structural parts of ordinary matter. Therefore, if it turns out that electron itself testifies in favor of the evrmion and its neutrino, the legality of the discussed procedure requires the elucidation of nature of a corresponding mechanism responsible for the steadiness of atomic system itself.

A given circumstance becomes more interesting if we now look at the following inequalities: 
\begin{equation}
m_{\epsilon}^{E}< m_{e}^{E}< m_{\mu}^{E}< m_{\tau}^{E},
\label{55}
\end{equation}
\begin{equation}
m_{\nu_{\epsilon}}^{E}< m_{\nu_{e}}^{E}<
m_{\nu_{\mu}}^{E}< m_{\nu_{\tau}}^{E},
\label{56}
\end{equation}
\begin{equation}
m_{\epsilon}^{W}> m_{e}^{W}> m_{\mu}^{W}> m_{\tau}^{W},
\label{57}
\end{equation}
\begin{equation}
m_{\nu_{\epsilon}}^{W}> m_{\nu_{e}}^{W}>
m_{\nu_{\mu}}^{W}> m_{\nu_{\tau}}^{W},
\label{58}
\end{equation}
which say that each of transitions (\ref{50}), (\ref{53}), and (\ref{54}) similarly to any 
of Eqs. (\ref{6})-(\ref{8}) cannot go at the expense of weak currents, and is a result of 
Coulomb interactions. On the other hand, these conditions do not contradict the fact that 
to each of the decays
\begin{equation}
\epsilon^{-}_{L,R}\rightarrow e^{-}_{L,R}{\bar \nu_{e R,L}}\nu_{\epsilon L,R}, \, \, \, \,
\epsilon^{+}_{R,L}\rightarrow e^{+}_{R,L}\nu_{e L,R}{\bar \nu_{\epsilon R,L}},
\label{59}
\end{equation}
\begin{equation}
\epsilon^{-}_{L,R}\rightarrow \mu^{-}_{L,R}{\bar \nu_{\mu R,L}}\nu_{\epsilon L,R}, \, \, \, \,
\epsilon^{+}_{R,L}\rightarrow \mu^{+}_{R,L}\nu_{\mu L,R}{\bar \nu_{\epsilon R,L}},
\label{60}
\end{equation}
\begin{equation}
\epsilon^{-}_{L,R}\rightarrow \tau^{-}_{L,R}{\bar \nu_{\tau R,L}}\nu_{\epsilon L,R}, \, \, \, \,
\epsilon^{+}_{R,L}\rightarrow \tau^{+}_{R,L}\nu_{\tau L,R}{\bar \nu_{\epsilon R,L}},
\label{61}
\end{equation}
\begin{equation}
e^{-}_{L,R}\rightarrow \mu^{-}_{L,R}{\bar \nu_{\mu R,L}}\nu_{e L,R}, \, \, \, \,
e^{+}_{R,L}\rightarrow \mu^{+}_{R,L}\nu_{\mu L,R}{\bar \nu_{e R,L}},
\label{62}
\end{equation}
\begin{equation}
e^{-}_{L,R}\rightarrow \tau^{-}_{L,R}{\bar \nu_{\tau R,L}}\nu_{e L,R}, \, \, \, \,
e^{+}_{R,L}\rightarrow \tau^{+}_{R,L}\nu_{\tau L,R}{\bar \nu_{e R,L}},
\label{63}
\end{equation}
\begin{equation}
\mu^{-}_{L,R}\rightarrow \tau^{-}_{L,R}{\bar \nu_{\tau R,L}}\nu_{\mu L,R}, \, \, \, \,
\mu^{+}_{R,L}\rightarrow \tau^{+}_{R,L}\nu_{\tau L,R}{\bar \nu_{\mu R,L}}
\label{64}
\end{equation}
respond the weak interactions. Of course, such transitions may be observed as the extremely fast processes, because parafermions (\ref{9})-(\ref{12}) and
\begin{equation}
(\tau_{L}^{-}, {\bar \nu_{\tau R}}), \, \, \, \,
(\tau_{R}^{-}, {\bar \nu_{\tau L}}),
\label{65}
\end{equation}
\begin{equation}
(\tau_{R}^{+}, \nu_{\tau L}), \, \, \, \,
(\tau_{L}^{+}, \nu_{\tau R})
\label{66}
\end{equation}
appear in them as a consequence of weak currents.

\vspace{0.8cm}
\noindent
{\bf 6. Leptons about the proton decay} 
\vspace{0.4cm}

We now remark that charge conservation in the processes
\begin{equation}
n^{-}_{L,R}\rightarrow p^{-}_{L,R}e^{+}_{R,L}\nu_{e L,R}, \, \, \, \,
n^{+}_{R,L}\rightarrow p^{+}_{R,L}e^{-}_{L,R}{\bar \nu_{e R,L}}
\label{67}
\end{equation}
assumed that the neutron and electronic neutrino possess the identical electric charges [35].

In the same way, one can reanalyze in the framework of the baryon [36] and lepton number 
conservation the new scheme for the neutron decay
\begin{equation}
n^{-}_{L,R}\rightarrow p^{-}_{L,R}\epsilon^{+}_{R,L}\nu_{\epsilon L,R}, \, \, \, \,
n^{+}_{R,L}\rightarrow p^{+}_{R,L}\epsilon^{-}_{L,R}{\bar \nu_{\epsilon R,L}}.
\label{68}
\end{equation}
It appears that the transitions (\ref{68}) are in favor of an equality of electric charges 
of the neutron and evrmionic neutrino.

Thus, Eqs. (\ref{67}) and (\ref{68}) would seem to indicate that either the neutron and neutrino are the electrically neutral or the hypothesis about the identicality of the electric charges of the proton, evrmion and electron is not valid. On the other hand, as follows from the flavor symmetry, a formation of the parafermions (\ref{9}), (\ref{10}), (\ref{51}), and (\ref{52}) can be explained by the availability in neutrino of a small charge [22]. At the same time, the existence itself of the neutron decay is by no means excluded experimentally [16]. At our sight, such a measurement can explain the unity of a neutrino universality principle in the systems of all fermions with the charge of evrmionic neutrino. This in turn implies that the fact [37] of lepton universality is a general and does not depend on the type of a particle with the charge of evrmion. 

If we start from the legality of a given procedure, using Eqs. (\ref{27}) and (\ref{45}) for the nucleons, we would establish the following equalities: 
\begin{equation}
(m_{\epsilon}^{K})^{2}=m_{l}^{E}m_{l}^{W}=
m_{p}^{E}m_{p}^{W},
\label{69}
\end{equation}
\begin{equation}
(m_{\nu_{\epsilon}}^{K})^{2}=m_{\nu_{l}}^{E}m_{\nu_{l}}^{W}=
m_{n}^{E}m_{n}^{W}.
\label{70}
\end{equation}

They together with Eqs. (\ref{24}), (\ref{42}) and the available data [16] in the literature
\begin{equation}
m_{p}^{E}=938.272\ {\rm MeV},
\label{71}
\end{equation}
\begin{equation}
m_{n}^{E}=939.565\ {\rm MeV}
\label{72}
\end{equation}
lead to laboratory estimates of nucleon weak masses 
\begin{equation}
m_{p}^{W}=2.8049552\cdot 10^{-5}\ {\rm eV},
\label{73}
\end{equation}
\begin{equation}
m_{n}^{W}< 5.6021883\cdot 10^{-18}\ {\rm eV}.
\label{74}
\end{equation}
Comparing their masses with lepton masses, it is not difficult to see that
\begin{equation}
m_{\epsilon}^{E}< m_{e}^{E}< m_{\mu}^{E}< m_{p}^{E}< m_{\tau}^{E},
\label{75}
\end{equation}
\begin{equation}
m_{\nu_{\epsilon}}^{E}< m_{\nu_{e}}^{E}<
m_{\nu_{\mu}}^{E}< m_{\nu_{\tau}}^{E}< m_{n}^{E},
\label{76}
\end{equation}
\begin{equation}
m_{\epsilon}^{W}> m_{e}^{W}> m_{\mu}^{W}> m_{p}^{W}> m_{\tau}^{W},
\label{77}
\end{equation}
\begin{equation}
m_{\nu_{\epsilon}}^{W}> m_{\nu_{e}}^{W}>
m_{\nu_{\mu}}^{W}> m_{\nu_{\tau}}^{W}> m_{n}^{W}.
\label{78}
\end{equation}

In these circumstances, any of the transitions (\ref{67}) and (\ref{68}) can carry out through the Coulomb interactions. But the difference in masses of fermions admits the weak decays of protons 
by the schemes
\begin{equation}
p^{-}_{L,R}\rightarrow n^{-}_{L,R}\tau^{-}_{L,R}{\bar \nu_{\tau R,L}}, \, \, \, \,
p^{+}_{R,L}\rightarrow n^{+}_{R,L}\tau^{+}_{R,L}\nu_{\tau L,R}.
\label{79}
\end{equation}

The nucleons of both types possess in addition the strong interaction, which is absent in leptons. However, as mentioned earlier [38], the neutrinos have with the field of emission the same strong interactions as the hadrons. Their existence in all types of leptons could refine the crossing point of spectra of masses of a different nature.

\vspace{0.8cm}
\noindent
{\bf 7. Conclusion}
\vspace{0.4cm}

The evrmion has the lowest electric mass within the four families of leptons and cannot decay by means of Coulomb interactions. Instead it possesses the large weak mass and almost always decays through the weak interactions.

The electron and muon are of those fermions in which the decay is carried out by the two ways 
either at the expense of the Coulomb or at the expense of the weak interaction.

Among the set of the studied fermions only the $\tau$-lepton and neutron have the extremely lower weak mass and must not decay by means of weak currents. However, their decay through the Coulomb interactions is not forbidden, since at such processes appears a crucial part of the electric mass. The weak mass of the proton is responsible for its transition into the neutron, $\tau$-lepton 
and neutrino with the lowest weak mass.

But in all these decays no masses are separately conserved [39] even at the account of their sign for the fermion and the antifermion. At the same time, neither of the investigated transitions is forbidden by any other conservation laws. Therefore, it is important to elucidate what is the
selection rule, which expresses the idea of mass conservation.

To solve this question, we will start from the fact that 
\begin{equation}
\sum_{l}e_{l}^{E}=const, \, \, \, \,
\sum_{\nu_{l}}e_{\nu_{l}}^{E}=const
\label{80}
\end{equation}
correspond in any $\beta$-decay to charge conservation.

Jointly with Eqs. (\ref{16}) and (\ref{32}), these selection rules are reduced to
\begin{equation}
\sum_{l}m_{l}^{E}m_{l}^{W}=const, \, \, \, \,
\sum_{\nu_{l}}m_{\nu_{l}}^{E}m_{\nu_{l}}^{W}=const.
\label{81}
\end{equation}

At first sight unlike the charge, no mass possesses the additivity properties. On the other hand, 
as we have already seen, each of charges satisfying the additivity conditions must be universal physical parameter. At the same time, a particle itself can possess simultaneously both the mass 
and the charge of the evrmion or its neutrino.

Taking into account this and that
\begin{equation}
\sum_{l}e_{\epsilon}^{K}=const, \, \, \, \,
\sum_{\nu_{l}}e_{\nu_{\epsilon}}^{K}=const,
\label{82}
\end{equation}
we cannot exclude the existence of an additive selection rule for the square of any of the universal masses of elementary particles
\begin{equation}
\sum_{l}(m_{\epsilon}^{K})^{2}=const, \, \, \, \,
\sum_{\nu_{l}}(m_{\nu_{\epsilon}}^{K})^{2}=const.
\label{83}
\end{equation}

Thus, nature itself unites the conservation laws of mass and charge in a unified whole. Thereby, 
it characterizes the behavior of electroweakly charged fermions both from the point of view of  Coulomb and from the point of view of weak parts of their mass. Therefore, each of all types of masses of any particle may serve as a criterion for a kind of scheme of its decay unless this is forbidden by unification laws.

\vspace{0.8cm}
\noindent
{\bf References}
\begin{enumerate}
\item
B. Pntecorvo, Zh. Eksp. Teor. Fiz. {\bf 6}, 429 (1958).
\item
Ya.B. Zel'dovich and M.Yu. Khlopov, Uspehi Fiz.
Nauk. {\bf 135}, 45 (1981).
\item 
R.S. Sharafiddinov, Bull. Am. Phys. Soc. {\bf 59}(5), T1.00004 (2014); e-print 

arXiv:hep-ph/0511065.
\item 
R.S. Sharafiddinov, Fizika {\bf B 16}, 1 (2007);
e-print arXiv:hep-ph/0512346.
\item 
A.K. Ciftci, R. Ciftci, H. D. Yildiz, and S. Sultansoy, Mod. Phys. 

Lett. {\bf A 23}, 1047 (2008).
\item 
A.K. Ciftci, R. Ciftci, and S. Sultansoy,
Phys. Rev. {\bf D 72}, 053006 (2005).
\item 
Ya.B. Zel'dovich, Dokl. Akad. Nauk SSSR. {\bf 91}, 1317 (1953).
\item 
E.J. Konopinsky and H. Mahmoud, Phys. Rev. {\bf 92}, 1045 (1953).
\item 
A. Van Der Schaaf, in Proc. Summer School on Particle Physics,
Zuoz, August 18-24, 2002 (Z\"urich, Switzerland, 2003), p. 269.
\item
P.A.M. Dirac, Proc. Roy. Soc. {\bf A 133}, 60 (1931).
\item 
P.A.M. Dirac, Phys. Rev. {\bf D 74}, 817 (1948).
\item 
M.A. Markov, Zh. Eksp. Teor. Fiz. {\bf 51}, 878 (1966).
\item
V.G. Kadyshevsky, M.D. Mateev, and M.V. Chizhov, Theor. Math. 

Phys. {\bf 45}, 1077 (1981).
\item
R.S. Sharafiddinov, Bull. Am. Phys. Soc. {\bf 59}(5), T1.00009 (2014); 
Spacetime Subst. {\bf 3}, 47 (2002); e-print arXiv:physics/0305008.
\item 
R.S. Sharafiddinov, Eur. Phys. J. Plus {\bf 126}, 40 (2011); e-print 

arXiv:0802.3736 [physics.gen-ph].
\item 
Particle Data Group, {\it Review of Particle Properties},
Phys. Rev. {\bf D 45} (1992).
\item 
J.H. Park, JHEP. {\bf 10}, 077 (2006); e-print arXiv:hep-ph/0607280.
\item 
O. Klein, Nature {\bf 161}, 897 (1948).
\item 
E. Clementel and G. Puppi, Nuovo Cimento {\bf 5}, 505 (1948).
\item 
J. Tiomno, J. Wheeler, Rev. Mod. Phys. {\bf 21}, 144 (1949).
\item 
T.D. Lee, M.H. Rosenbluth, and C.N. Yang, Phys. Rev.
{\bf 75}, 905 (1949).
\item 
R.S. Sharafiddinov, Can. J. Phys. {\bf 92,} 1262 (2014); e-print 

arXiv:0807.3805 [physics.gen-ph].
\item 
J.L. Lucio Martinez, A. Rosado, and A. Zepeda,
Phys. Rev. {\bf D 29}, 1539 (1984).
\item 
M. Dvornikov and A. Studenikin, Phys. Rev. {\bf D 69}, 073001 (2004); e-print 

arXiv:hep-ph/0305206.
\item 
R.S. Sharafiddinov, Phys. Essays {\bf 19}, 58 (2006); e-print arXiv:hep-ph/0407262.
\item 
R.S. Sharafiddinov,  Bull. Am. Phys. Soc. 59(5), T1.00005 (2014);
Spacetime Subst. {\bf 3}, 132 (2002); e-print arXiv:physics/0305014.
\item 
I.S. Batkin and M.K. Sundaresan, J. Phys. {\bf G 20}, 1749 (1994).
\item 
S.L. Glashow, Nucl. Phys. {\bf 22}, 579 (1961).
\item 
A. Salam, J.C. Ward, Phys. Lett. {\bf 13}, 168 (1964).
\item 
S. Weinberg, Phys. Rev. Lett. {\bf 19}, 1264 (1967).
\item 
J. Bernstein, M. Ruderman, and G. Feinberg, Phys. Rev. {\bf 132}, 1227 (1963).
\item 
G.G. Raffelt, Phys. Rep. {\bf 320}, 319 (1999).
\item 
F.J. Browning, D. Chang, and W.Y. Keung, Phys. Lett. {\bf B 444}, 142 (1998).
\item 
M. Jezabek and P. Urban, Acta Phys. Polon. {\bf B 30}, 3353 (1999).
\item  
R.S. Sharafiddinov, Int. J. Theor. Phys. {\bf 55}, 3040 (2016); Bull. Am. Phys. 
Soc. {\bf 57}(16), KA.00069 (2012); e-print arXiv:1004.0997 [hep-ph].
\item 
Ya.B. Zel'dovich, Dokl. Akad. Nauk SSSR. {\bf 86}, 505 (1952).
\item 
A. Pich, CERN Cour. {\bf 40}, 20 (2000).
\item 
L.B. Okun, Yad. Phys. {\bf 45}, 158 (1987).
\item 
L.B. Okun, Uspehi Fiz. Nauk. {\bf 158}, 511 (1989).
\end{enumerate}
\end{document}